\documentclass[11pt,notitlepage]{article}
\usepackage{amsmath,amssymb, amsthm, enumerate,boxedminipage}

\usepackage{todonotes}
\usepackage{xspace}
\usepackage{url}
\usepackage{fullpage}

\usepackage{amsmath,amssymb, amsthm, enumerate,boxedminipage}

\usepackage{algorithmic}
\usepackage{graphicx}
\usepackage{textcomp}
\usepackage{xcolor}
\usepackage{lipsum}

\usepackage[T1]{fontenc} 
\usepackage{wrapfig,floatrow}
\usepackage{subcaption}
\usepackage{amsthm}
\usepackage{algorithmic}
\usepackage{graphicx}
\usepackage{textcomp}
\usepackage{xcolor}
\usepackage{tcolorbox}
\usepackage{array}
\usepackage{tabulary}
\usepackage{xspace}
\usepackage{colortbl}
\usepackage{enumitem}
\usepackage[normalem]{ulem} 
\usepackage{url}
\DeclareCaptionFont{tiny}{\tiny}
\usepackage{balance}
\usepackage{ctable} 

\usepackage[ruled]{algorithm2e}
\usepackage{xpatch}

\usepackage{hyperref}
\def\BibTeX{{\rm B\kern-.05em{\sc i\kern-.025em b}\kern-.08em
    T\kern-.1667em\lower.7ex\hbox{E}\kern-.125emX}}

\newif\ifcomments
\commentstrue

\definecolor{darkblue}{rgb}{0.0,0.0,0.75}

\def\mval{-4pt}
\def\lval{-30pt}

\begin{document}

\title{\vspace{\lval}Correction to ``Byzantine Agreement in Expected Polynomial Time"}

\author{Valerie King \thanks{val@cs.uvic.ca; Department of Computer Science, University of Victoria, P.O.
Box 3055, Victoria, BC, Canada V8W 3P6.} \and Jared Saia \thanks{saia@cs.unm.edu; Department of Computer Science, University of New Mexico, Albuquerque, NM 87131-1386.}}

\date{}

\maketitle

\begin{abstract}
	This is a correction to the paper ``Byzantine Agreement in Expected
Polynomial Time" which appeared in the Journal of the ACM in 2016. It corrects a failure in the paper to consider the adversary's ability to decide the number of fair coinflips in an iteration, where this number ranges between $n(n-t)$ and $n(n-2t)$. \vspace{\mval} 
\end{abstract}


\section{Introduction}\vspace{\mval}

In~\cite{DBLP:journals/jacm/KingS16}, King and Saia describe an algorithm to solve Byzantine agreement in expected polynomial time. Their general strategy is to run an algorithm based on work of Ben-Or~\cite{ben1983another} (cf. Bracha~\cite{DBLP:journals/iandc/Bracha87}) that depends on simulating a global coin.  In every iteration of the Ben-or type algorithm, there is a direction that is \emph{good} in the sense that if the simulated global coin lands in that direction, the processors come to agreement and terminate.

 King and Saia simulate a global coin via summing coinflips from all processors, and measuring the deviation from zero of this sum.  They show that in each iteration, with constant probability there is large ``good" deviation by the good coinflips in the good direction.  When this occurs, if there is no agreement,  then there must be ``bad" deviation caused by the adversarially-controlled (bad) processors which counterbalances the good deviation. Eventually, the deviation by bad processors becomes unusually frequent and detectable, so that good processor will eventually exclude these from their calculations in future global coin simulations. 


In each iteration of the Ben-Or type algorithm, each good processor generates a stream of up to $n$ fair coinflips.  
In Section 4.1, Property 2 of x-sync ensures that at most $t$ of the coinflip streams generated by good processors  contain fewer than $n$ coinflips.  It was recently observed by Melnyk, Wang and Wattenhoffer~\cite{MWW18} that the analysis in~\cite{DBLP:journals/jacm/KingS16} failed to take into account the effect that an adversarial stopping strategy on these $t$ streams may have on the value of the sum of coinflips generated by good processors.

In this brief announcement, we correct this issue.  First, we show (Fact 3 below) that we can bound the maximum value obtained during a stream of $tn$ coins, and pessimistically use this maximum in lieu of the final sum of $tn$ coins.  Next,  we describe additional slight changes required in the analysis of the remainder of the paper to account for this new bound.  This includes decreasing the constant factor in the resilience for the polynomial time algorithm, Variant 1.\vspace{\mval}




\section{Corrections in Section 5}

First, insert the following result~\cite{Zitkovic}, as Fact 3 in Section 5.1. \\

\noindent
{\bf Fact 3:} Let $M_n$ be the maximum value of a symmetric, $1$-dimensional random walk after $n$ steps and let $S_n$ be the value of the random walk at step $n$.  For a symmetric random walk, for any $r \geq 1$, $Pr(M_n \geq r)= Pr(S_n = r) + 2 Pr(S_n > r)$, which implies     
$Pr(M_n \geq  r) < 2 Pr(S_n \geq r)$. \medskip

The JACM paper identifies two sources of bad deviation other than that caused by coinflips from bad processors:  (1) coinflips from $\leq t$  good processors which have been excluded because they are thought to be bad which sum to $\leq \beta/2$; and (2) $t$ deviation caused by a total of $t$ ``ambiguous"  coins (from the asynchrony).  In this writeup, we add (3), the effect of the adversary possibly stopping up to $t$ of the good coinflip streams.

\subsection{Our Corrections}
 We change $\beta/2$ to $\beta/4$ for each of the coinflips of type (1)  and type (3) and modify Lemma 5.2 to account for these effects. \\

\noindent
{\bf Lemma 5.2. }\ (1)  A stream of between $1$ and $nt$ coin flips which can be stopped adversarially anywhere in this range has deviation exceeding $\beta/4 = \sqrt{2n(n-t)}/4 -t/2$ with probability at most $(1/2) e^{-(\beta/4)^2}/2nt$. If $t < .005n$, then this probability is at most $e^{-11}$.\\
(2) A stream of between $n(n-2t)$ and $n(n-t)$ coin flips which can be stopped adversarially anywhere in this range has deviation of at least $\alpha'=\sqrt{2n(n-2t)} -\beta/4$ in any specified direction with probability exceeding $1/20$. 

\begin{proof}
(1) Consider a random walk of $nt$ steps, starting at the origin, where each step is equally likely to be 1 or -1. By Fact 3, the probability that its maximum value is at least $\beta/4$ is no more than twice the probability that the sum of $nt$ coin flips is at least $\beta/4$, which is at most $2 e^{-(\beta/4 )^2/2tn}$ (by Lemma 5.2 old). If $t< .005n$, then this probability is at most $e^{-11}$. 

\noindent
(2) The stream consists of the first $n(n-2t)$ good coins plus a stream which the adversary can stop at any point afterwords. By Lemma 5.1, the first part of the stream has deviation $\alpha$ with  probability greater than $.211$. The second part of the stream can, in the worst case, be adjusted by the adversary to the maximum deviation achieved which equals or exceeds $\beta/4$ (in the opposite direction) with probability at most $ e^{-11}$. The probability that the entire stream has deviation at least $\alpha'$ is at least the probability that the first event occurs and the second doesn't. 
Hence the probability that the entire stream has deviation $\alpha'$ is at least $.211-e^{-11}>1/20$.
\end{proof}

\noindent
\medskip
Additional corrections in Section 5 are as follows.
\begin{itemize}
 \item In Lemma 5.3: (1)  Replace $\alpha$ by $\alpha'$; and (2) Replace $\beta/2$ by $\beta/4$.
 \item In proof of Lemma 5.3: Change ``From Lemma 5.1" to  ``From Lemma 5.2(2)"
\item In Lemma 5.4: (1) Change $\alpha$ to $\alpha'$; and (2) Change $\beta/2$ to $\beta/4$
\item In proof of Lemma 5.4, line 7 page 14: Change $\beta/2$ to $\beta/4$
\item In Lemma 5.5: Change $t< n/36$ to $t< 1/72$.
\end{itemize}

\section{Corrections for Variant 1}
For Variant 1: 
Let $H$ be the matrix filled by a stream of independent, random ${-1,1}$ coinflips except in up to $t$ adversarially chosen columns.  In these columns, the adversary picks points to stop the stream  as it is filling a column, and the rest of the column (the suffix) is filled with 0's. 

Then $H =  H' + W$ where $H'$ is a matrix of ${-1,1}$ independent, random coin flips, and $W$ is all 0's except the suffixes in columns changed by the adversary, where its entries are the negation of the columns of $H'$.

Now change the definition of $G$, in Section 6, second paragraph after Theorem 6.1 as follows.  ``All entries of $G$ in columns controlled by bad processors are set to $0$.  For $j$ corresponding to a good processor, the $(i,j)$ entry of $G$ is the sum of the $j$-th column for the matrix $H$ generated in iteration $i$ of Algorithm 5."

Then insert the following two paragraphs immediately before Corollary 6.2.  ``Note that $G=R+Z$, where $R$ is a matrix where every entry in each good column is set to the sum of $n$ fair coinflips, and every entry in each bad column is set to $0$.  Additionally, $Z$ is a matrix where the $(i,j)$ entry is the sum of the $j$-th column of the matrix $W$ generated in iteration $i$ of Algorithm 5."

``By a property of matrix norms, $|G| \leq |R| +|Z|$.
Hence,  $Pr(|G| >a)  \leq  Pr (|R|+|Z| >a) \leq Pr(|(R|>a/2) + Pr(|Z|>a/2)$, where the last step holds by a union bound. 
$Pr(|R|>a/2)$ is bounded by Theorem 6.1.  $Pr (|Z| > a/2)$ is bounded by Theorem 6.1 as  well. Note that $Z$ is a matrix of independent random variables  some of which are set to $0$ and some of which are distributed as the sum of a number of coinflips  between $1$ and $n$.  The number is set adversarially but this does not affect the analysis. 

Next, change the statement of Corollary 6.2 to refer to both the matrices $Z$ and $R$, instead of $G$, and insert the following statement after Corollary 6.2:  
``Applying Corollary 6.2 to both $Z$ and $R$,
we have $Pr(|G| > (6+ 2\epsilon)\sqrt{n(m+n)}) < 2(m+n)^{-1}$." 

In Lemmas 6.3, 6.4, 6.5, 6.6, 6.7, and their proofs, replace the expression $3+\epsilon$ with $6+2\epsilon$, and the expression $4+\epsilon$ with $7+2\epsilon$. 
In proof of Lemma 6.7, replace $1/(m+n)$ by $2/(m+n)$ each time it appears. 
In statement of Lemma 6.7 and Theorem 1.1, replace $t < 4.25*10^{-7} n$ by $t<3.3 * 10^{-8}n$. \medskip

In the paragraph before Lemma 6.4, replace  "When $t< n/36$, $\beta/2 > 23n/36$"  with "When $t< 10^{-6}n$, $(\beta/2)^2 > .49999n^2$" and 
replace the sentence ``Then this inequality holds..." with  ``Then this inequality holds when $t < \frac{c_1n(.183^2)\beta^2}{6(7+2\epsilon)^2n^2}$, which requires $t < (2/3) (.001) (.0183)^2 (.49999)^2 (7+2 \epsilon)^{-2} n \leq 1.14 * 10^{-9}n$.

\section{Corrections for Variant 2}
For Variant 2, 
we need to change Lemma 7.1 as follows.\\  Bottom of page 18: Omit ``Further, note...Throughout the proof".  Insert ``The maximum sum of the coinflips of $t$ possibly incomplete streams of up to a total of $c_1 m n t$ coins is no greater than the maximum value of a random walk reached during this many steps. Hence, by Lemma 5.2,  the probability that $X \geq (\beta/6)c_1m$ is no greater than $2 Pr(Y\geq (\beta/6)c_1m)$, where $Y$ is the summation of $ntc_1m$ random coinflips." \medskip

\noindent
Make the following changes to the proof of Lemma 7.1:

Substitute $Y$ for $X$ in the third paragraph of Lemma 7.1 and the first line of the first set of equations. In the second set of equations, add a factor of 2 in front of the right side of each equation except the last.

{\small
\bibliographystyle{abbrv}
\bibliography{jam.bib}

}
\end{document}